\documentclass[aps,preprint,amssymb,12pt,floatfix]{revtex4}
\usepackage{subfigure}
\usepackage{graphicx}
\usepackage{rotating}
\usepackage{tabularx}

\begin{document}

\title{Proteins associated with diseases show enhanced sequence correlation between charged residues}

\renewcommand{\thefootnote}{\fnsymbol{footnote}}
\author{Ruxandra I. Dima and D. Thirumalai\footnote{Corresponding author:
Institute for Physical Science and Technology, University of Maryland,
College Park, MD 20742,
phone: 301-405-4803; fax: 301-314-9404; e-mail: thirum@Glue.umd.edu}}
\affiliation{\em \small Institute for Physical Science and Technology\\ 
        University of Maryland, College Park, MD 20742\\
 (accepted for {\em Bioinformatics})}

\date{\small \today}

\maketitle

\baselineskip = 22pt


\textbf{\large Abstract} 

\noindent
{\bf \em Motivation}: Function of proteins or a network of interacting
proteins often involves communication between residues that are well
separated in sequence. The classic example is the participation of distant residues
in allosteric regulation. Bioinformatic and structural analysis methods have been
introduced to infer residues that are correlated. Recently, increasing attention 
has been paid to obtain the sequence properties that determine the tendency of
disease related proteins (A$\beta$ peptides, prion proteins, transthyretin etc.)
to aggregate and form fibrils. Motivated in part by the need to identify sequence
characteristics that indicate a tendency to aggregate, 
we introduce a general method that probes covariations in
charged residues along the sequence in a given protein family.  The
method, which involves computing the Sequence Correlation Entropy (SCE) using
the quenched probability $P_{s_k}(i,j)$ of finding a residue pair at a given
sequence separation $s_k$, allows us to classify protein families in terms of their SCE. 
Our general approach may be a useful way in obtaining evolutionary 
covariations of amino acid residues on a genome wide level. 

\noindent
{\bf \em Results}: We use a combination of SCE and clustering based on the
principle component analysis to classify the protein families. From an
analysis of 839 families, covering approximately 500,000 sequences, we find that proteins with
relatively low values of SCE are predominantly associated with various
diseases. In several families, residues that give rise to 
peaks in $P_{s_k}(i,j)$ are clustered in the three dimensional structure.
For the class of proteins with low SCE values there are
significant numbers of mixed charged-hydrophobic $(CH)$ and charged-polar
$(CP)$ runs. Our findings suggest that low values of SCE and the presence of $(CH)$
and/or $(CP)$ may be indicative of disease association or tendency to aggregate.
Our results lead to the hypothesis that functions of proteins with similar SCE
values may be linked. The hypothesis is validated with a few anecdotal examples.
The present results also lead to the prediction that the overall charge correlations in
proteins affect the kinetics of amyloid formation -- a feature that is common to all
proteins implicated in neurodegenerative diseases.



\vspace{1cm}

{\bf \large Introduction}

The classification of
proteins based on their structures into families is useful not only in assigning distinct functions but 
also for examining the evolution of sequences with related functions. 
Because proteins in a family are descendants of the same ancestral protein, it is logical to
postulate that the observed sequence differences are the result of evolutionary pressure which vary greatly 
across distinct organisms. Sequence variations are tempered by the requirements of native state stability
and function. 
Destabilization of the folded state by mutations at one site can be compensated by mutations at distant or
nearby sites (Lesk and Chothia, 1980; Neher, 1994). Thus, it is important to study covariations of amino acids at
distinct sites to decipher if there
is communication between the two, especially as it pertains to function. Long-range communications
between several residues
(both along the sequence and across domains or interfaces in protein complexes) are crucial
for biological function. 
Thus, it might also be necessary to introduce methods to infer multi-site variations across sequences in
order to understand a number of issues in proteomics.

Correlations between amino acids in protein families have been probed using computational methods
beginning with the classic works of Lesk and Chothia
and Altschuh et al. (Lesk and Chothia, 1980; Altschuh et al. 1987). 
Several studies (Neher, 1994; Taylor and Hatrick, 1994; Pollock and Taylor, 1997)
have discovered
relationships between coordinated amino acid changes that occur during evolution
and the corresponding structural alterations. The working
hypothesis in these studies is that a mutation at a site that compromises 
the function is often compensated by a mutation at another site that is in proximity
in the three dimensional structure. The difficulty
in validating the working hypothesis arises largely because
multi-correlation effects, which are difficult to capture (Pollock and Taylor, 1997), 
can be important in compensating a given
mutation. Nevertheless, the computational methods that capture sequence covariations 
have provided insights in a number of areas of protein science
(Landgraf et al., 2001; Pazos et al., 1997; Olmea et al., 1999; Fariselli and Casadio, 1999; Lockless and Ranganathan, 1999; Lichtarge and Sowa, 2002). 

To infer the functional importance of correlated mutations, it is crucial to include physico-chemical
characteristics of amino acids (charge, volume of side-chains, hydrophobicity, etc.) 
to describe the positions in a multiple 
sequence alignment (MSA) (Lesk and Chothia, 1980; Neher, 1994).
Based on a study of divergent evolution in a set of protein families with known folds (Chelvanayagam et al., 1997) 
it has been argued that only charged residues show discernible
covariation at all evolutionary distances. 
With these observations in mind, we have investigated, using a new method, covariations among charged residues 
in 839 families. To obtain such correlations  
we introduce a function, the Sequence Correlation Entropy (SCE), that measures the extent
to which {\em two sites along a given sequence} are coupled. The values of SCE for protein families show that 
families/functions are associated with specific patterns of charges. There is a relationship between the degree of 
correlation of charged amino acids and the disease associations of 
a family. Families with high degree of correlation between charged residues
also have many significant mixed charged-hydrophobic/polar runs in the sequences. These
significant findings suggest that charges occur
in well defined patterns. Furthermore, variations in charges along sequences occur 
often in a correlated fashion in the evolutionary process.\\


{\bf \large System and Methods}\\

{\bf Sequence correlation function and the associated ``entropy'':}
We introduce a general measure that probes correlation between specific
residues that are separated by a given length for a database of sequences. Here
we focus on charged residues (D and E are negatively charged, and K and R are positively charged). 
To measure the correlation along the sequence between two charged residues,
$i$ and $j$ ($i,j \;\;\; \epsilon \;\;\; \{$+,-$\}$), we introduce the Sequence Correlation Entropy
(SCE) 

\begin{equation}
S(i,j) = - \sum_{k=1}^{I_{max}(i,j)} P_{s_k}(i,j) ln(P_{s_k}(i,j))
\label{eqn:S_ij}
\end{equation}

\noindent
where $s_k$ is the sequence separation between the residues, $k$ labels the pairs ($i,j$),
and $I_{max}(i,j)$ is the total number of sequence 
separations between residues $i$ and $j$ along the sequences of the family. 
We choose those pairs for which 
the locations of $i$ and $j$ are consecutive, or only those $(i,j)$ pairs for which there 
is no identical pair located between them. 
The probability of finding residues $i$ and $j$ at $s_k$ in the MSA is

\begin{equation}
P_{s_k}(i,j) = \frac{1}{n_{seq}(i,j)} \sum_{l=1}^{n_{seq}(i,j)} \frac{n^{(l)}(i,j)[s_k]}{n^{(l)}(i,j)}
\label{eqn:Pk_ij}
\end{equation}

\noindent
where $n_{seq}(i,j)$ is the number of sequences in the MSA, $n^{(l)}(i,j)$ is the number of pairs
of the type $(i,j)$ in sequence $l$ and $n^{(l)}(i,j)[s_k]$ is the number of
pairs of type $(i,j)$ from sequence $l$ at separation $s_k$. This equation is meaningful only if
the statistical ensemble  contains at least one pair of type $(i,j)$.
Note that $P_{s_k}(i,j)$ satisfies the normalization condition
$\sum_{k=1}^{I_{max}(i,j)} P_{s_k}(i,j) = 1$. Because the SCE uses a ``quenched'' sum 
(no preaveraging over all the sequences in the MSA) over the sequences of a given family, 
significant correlations, if present, can be gleaned. 
In contrast, in the mutual information function the equivalent of Eqn.(\ref{eqn:Pk_ij}) would be 
\begin{equation}
\frac{\sum_{l=1}^{n_{seq}(i,j)} n^{(l)}(i,j)[s_k]}{\sum_{l=1}^{n_{seq}(i,j)} 1}
\end{equation}
which involves implicit preaveraging over all the sequences 
in the MSA and might therefore mask correlations between residues. 
The SCE depends only
on $s_k$ and $P_{s_k}(i,j)$ regardless of where $i$ and $j$ are located. 
This accounts for the possibility that, to preserve the function, small 
rearrangements in the sequence could be preferred to compensatory changes 
at fixed positions along the sequence.

To assess the significance of SCE it is necessary to
calculate its expected value if the pairs are randomly distributed. For a protein family, with 
$L_{FAM}$ being the sequence length in the MSA, let $N(i,j)$ be the total number of pairs of type
($i,j$) and $I_{max}(i,j) = L_{FAM} -1$ be the maximum sequence separation between pairs of residues. 
If the ($i,j$) pair occurs randomly, the expected value of $P_{s_k}^{(rand)}(i,j)$ is
$P_{s_k}^{(rand)}(i,j) = \frac{1}{I_{max}(i,j)}$ provided $N(i,j) \geq I_{max}(i,j)$.
Otherwise, $P_{s_k}^{(rand)}(i,j) = \frac{1}{N(i,j)}$.
To compare results for SCE from all families on equal footing, we use

\begin{equation}
\bar{S}(i,j) = \frac{S(i,j)}{S^{(rand)}(i,j)} \times 100
\label{eqn:ratio_Sij}
\end{equation}

\noindent
where $S^{(rand)}(i,j) = - ln(P_{s_k}^{(rand)}(i,j))$.
If $\bar{S}(i,j)$ = 0 it implies
that amino acids $i$ and $j$ always occur at separation $s_k$ in all members of
the family. A relatively small value of $\bar{S}(i,j)$ means that there is a preferred
sequence separation $s_k$ for the ($i,j$) pair. \\

{\bf Significant mixed runs of charged and hydrophobic/polar residues:}
Karlin et al. (Karlin, 1995; Karlin et al., 2002) found that,
in five eukaryotic genomes, multiple long runs of given types of 
amino acids occur in proteins associated with diseases. 
For example, multiple long runs of glutamine, alanine, and serine
dominate in {\em Drosophila melanogaster} whereas in human sequences a preponderance of glutamate, proline,
and leucine is found.
Guided by these findings, we searched for
significant mixed charged-hydrophobic (CH) or charged-polar (CP)
runs. A mixed CH (CP) run is the longest possible segment of consecutive 
amino acids along the sequence such that the first and the last positions are occupied by charged 
amino acids while residues in between are either charged or hydrophobic (polar). If
$P_{random} = (P_{+})^{n_{+}} (P_{-})^{n_{-}} (P_{H})^{n_{H}} L_{seq} < 10^{-3}$
where $P_{+}$, $P_{-}$, $P_{H}$ are the percentage of +, - charged, hydrophobic (H) 
residues in the whole sequence, $n_{+}$, $n_{-}$ and $n_{H}$ are the numbers of each such 
type of residue in the given run and $L_{seq}$ is the length of the sequence, then a CH run is significant.
Significance for (CP) runs is similarly defined.
Using the number of significant mixed CH runs ($N_{run}(CH)$) and CP runs ($N_{run}(CP)$) in each sequence
in the MSA whose real length is at least half of the length of the alignment, we calculated
the average number of significant mixed runs per sequence
$R_{run} = \frac{N_{run}}{N_{seq}}$; $N_{run}$ is either $N_{run}(CH)$ or $N_{run}(CP)$, and $N_{seq}$ is the number
of sequences in the family. \\

{\bf Database of aligned sequences:}
We used two types of sequence alignments: (i) One obtained from the Pfam database 
(Bateman et al, 2002) (http://www.sanger.ac.uk/Software/Pfam/index.shtml) which is composed of 
families of protein domains; (ii) The other is produced by
aligning a query sequence with similar sequences using the 
Psi-Blast software (Altschul et al, 1990) (http://www.ncbi.nlm.nih.gov/BLAST)
with the default settings. In the February 2002 release Pfam had 3360 
protein families that covered 68\% of protein sequences.
For some families (prions and globins) we used both the Pfam database and the 
Blast alignment to check whether the two approaches lead to comparable 
results. For other sequences ($\tau$ protein or Sup35), for which
there is no known 3D structure, we used only the Blast alignment.

To compute SCE, we considered a dataset of 839 families from Pfam (the list is available upon request). 
The criteria for choosing
the families were: i) the average length of sequences in the 
family be at least 40 residues; ii) the number of members in
the family be large so that an ensemble of ($i,j$) pairs can be created; 
iii) the families give a good coverage of the
various protein functions. The database of families (``All'') consists of 
heat shock proteins (21, ``HSP''), nucleic acids (DNA or RNA) binding proteins (152, ``NA''), 
disease-related proteins (prions, other amyloidogenic proteins, cancer, allergens, 
toxins) (40, ``Disease''), viral proteins (209, ``Viruses'') including viral nucleocapsid 
proteins (26, ``Capsid'') and ``normal'' proteins (595, ``Normal''). 
The number of families is given in parenthesis. For example, the families from the ``Disease'' class represent
the subset of all the families retrieved from Pfam with keyword ``disease'' which satisfy also the above
mentioned criteria for statistical significance. 
The functions of the families are diverse enough that we can draw meaningful conclusions.\\

{\bf \large Implementation}\\
 
\begin{figure}[htbp]
        \subfigure[]{
        \label{fig:EV4_EV1}
        \includegraphics[width=4.0in,height=2.5in]{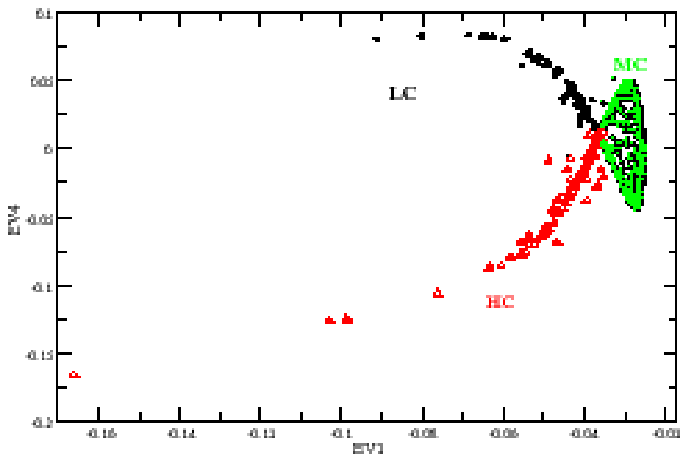}}
        \vspace{5mm}\\
        \subfigure[]{ \label{fig:EV4_EV2}
        \includegraphics[width=4.0in,height=2.5in]{EV4_EV2.eps}}
\caption{}
\label{fig:PCA_on_SCE}
\end{figure}

{\bf Disease associations based on clustering of sequence correlation entropy:}
The $P[\bar{S}(i,j)]$ distributions for the (+,+), (+,-) and (-,-) are broad
(Fig. \ref{fig:S_pair(C1,C2)_gaps}). Therefore any attempt to classify families based
entirely on these distributions is bound to be arbitrary. A more reliable method is to 
use a clustering procedure to divide the families according to their $\bar{S}(i,j)$ values.
We start by constructing a 839 $\times$ 3 matrix with the rows representing the families and 
the columns corresponding to $\bar{S}(i,j)$ values.
Inspired by the Principal Component Analysis (PCA) clustering procedure (Jolliffe, 1986),
we transformed the above matrix into the 839 $\times$ 839 matrix of the euclidean distances 
between all pairs of families. An analysis of the eigenvalues of this matrix shows
that the first 4-5 eigenvalues are much larger in magnitude than the others.
Therefore, if there exists a tendency of the protein families to cluster then such a tendency 
will manifest itself in the behavior of the eigenvectors associated to the largest eigenvalues
(because higher order eigenvectors are bound to remove structure from the data points).
Indeed, the plot of the second eigenvector (EV2) versus the first eigenvector (EV1) (data not shown) 
reveals two clusters of data: one corresponding to positive values of EV2, the other 
corresponding to negative values of EV2. But the boundary between the two clusters is not well
defined. A much better picture is provided by the plots of EV4 versus EV1 from Fig.(\ref{fig:EV4_EV1})
and EV4 versus EV2 from Fig.(\ref{fig:EV4_EV2}).
Both graphs present three regions which we represent by filled triangles, filled circles and stars. 
A mapping of the points from one graph to the other shows that the corresponding regions are populated by
almost the same set of points. But because the number of points in the corresponding regions from the
two graphs varies somewhat, we choose to define the
clusters based on Fig.(\ref{fig:EV4_EV2}). This choice leads to a more balanced
distribution of points in each cluster: the {\bf HC} cluster contains 210 points, the {\bf MC} cluster 
has 361 points, and the remaining 268 points are in the {\bf LC} cluster.
By mapping the points from the three clusters to families and their $\bar{S}(i,j)$ values, 
we can therefore classify the protein families in 3 classes:    
(1) Highly Correlated ({\bf HC}) families have at least two of the pairs satisfy the
constraints $\bar{S}(+,+) \leq 52\%$, $\bar{S}(+,-) \leq 42\%$, and
$\bar{S}(-,-) \leq 50\%$. (2) If at least two of the pairs satisfy 
$\bar{S}(+,+) = (53\%-63\%)$, 
$\bar{S}(+,-) = (43\%-54\%)$ and $\bar{S}(-,-) = (51\%-60\%)$ then the family
is considered to have moderate correlation ({\bf MC}). (3) When at least
two of the pairs lie outside the range, we assume that there is little correlation ({\bf LC}).

Many protein families known to be associated with various diseases belong to the 
{\bf HC} class (Table \ref{table:correl_charged_gaps}).
Examples of families belonging to the {\bf HC} class are prions,
A$\beta$ peptide, the $\tau$ protein and Sup35 (one of the {\em Yeast} prions) 
which are all known to aggregate and form fibrils. This result correlates well
with the findings of various studies of protein aggregation that for prions 
charged residues play a key role (Billeter et al., 1995). Our prediction supports
the observation that
mutations of charged residues drastically affect the fibrillization kinetics
in a variety of proteins in which aggregation occurs from the unfolded state 
(Massi et al., 2002; Chiti et al., 2003).   
Other families with high degree of sequence correlation between charged amino
acids represent proteins that cause diseases like viruses 
(HCV, Adeno-hexon, Vpu, Gag-p17), Androgen receptor (Kennedy disease), 
the lyme disease protein and P53 (whose malfunctioning is linked to cancer). 
Proteins which bind nucleic acids (DNA and RNA which are highly charged) 
like DNA-polB, recA and IF3, together
with proteins that are implicated in the response of the organism to environmental 
stress (PAL). 
The largest category of proteins represented 
in the {\bf HC} class are those associated with various diseases (Table \ref{table:results_all_fam}). 
Families that belong to the {\bf MC} class represent a mixture of 
structural proteins, enzymes, transport proteins and some disease-related proteins.
Some examples are: Aerolysin (related to deep wound infections), aldedh (allergens),
Alpha\_TIF (viral protein), Clathrin (structural protein in vesicles), FMN\_dh
(electron transfer), and GCV\_H (glycine cleavage H-protein).

Three control calculations are needed to ascertain if the classification of protein
families based on the clustering of SCE is reliable. 

(a) It is likely that the high sequence correlation
among charged residues can arise because of unusual sequence composition in disease
related proteins. If the fraction of total number 
of charged residues in a sequence exceeds the typical 23\% observed in protein structures
(Creighton, 1993), then
one might expect it to belong to the {\bf HC} class. Computation of the sequence composition of charged residues in 
all 839 families shows no correlation between the observed fraction of charged residues  
and the associated value of $\bar{S}(i,j)$ (RID, unpublished).

(b) To ascertain if our findings are a result of high
sequence identity, we explored the relationship between the average sequence identity
in a family (as presented in the Pfam entry) and its class based on $\bar{S}(i,j)$ values.
The distributions of sequence identities for families belonging 
to the three classes show that families with similarities above 90\%
belong to the {\bf HC} class, while families with similarities below 15\% are
most likely to belong to the {\bf LC} class. This is exactly what one expects based on
the Eqn.(\ref{eqn:S_ij}) and Eqn.(\ref{eqn:Pk_ij}). But, in general, there is no
good correlation between the sequence similarity in a family and its class (Fig.\ref{fig:Class_family_Seqid}).
There is considerable variation in the sequence identities between families
in both the {\bf HC} and {\bf MC} class (Fig.\ref{fig:Class_family_Seqid}).
Both these control calculations show that the values of SCE among charged residues
may indeed be associated with the function of the protein.\\

(c) To determine the significance of the $\bar{S}(i,j)$ values in the various families, 
a comparative analysis with a random dataset is required. For this, we built 100,000 sets each 
containing 1000 sequences of length 100. Each position in a sequence was assigned one of the 20 types of amino acids
with equal probability. The corresponding distributions of $\bar{S}(i,j)$ values (data not shown) are narrow with 
averages corresponding to 69\%, 57\% and 76\% for $\bar{S}(+,+)$, $\bar{S}(+,-)$, and $\bar{S}(-,-)$
respectively. Using these data sets, the {\em Pvalues for $\bar{S}(i,j)$ in the {\bf HC} class 
are $< 10^{-5}$, which shows that the calculated $\bar{S}(i,j)$ values are very significant}. 
More importantly, the link between $\bar{S}(i,j)$ and the tendency to aggregate is indeed
meaningful.\\

{\bf Specific sequence separations in charged residues may be preferred in proteins belonging to the HC class:}
Plots of $P_{s_k}(i,j)$ as a function of of $s_k$ for three of the families in the 
{\bf HC} class show (Figs.\ref{fig:P_prions_gaps}) 
that there is considerable structure in $P_{s_k}(i,j)$ which implies that in these families
there is distinct correlation among charged residues at preferred values of $s_k$. 
Surprisingly, despite the similarity in the overall degree of correlation in prions and 
DNA-polB and HCV capsid proteins, the behavior of 
$P_{s_k}(i,j)$ in prions resembles more closely that of the HCV capsid: in both cases
there are a few peaks separated by deep valleys. On the other hand, in DNA-polB 
$P_{s_k}(i,j)$ decays smoothly with the increase in $s_k$. 

\begin{figure}[htbp]
        \subfigure[]{
        \label{fig:1qlx_face1}
        \includegraphics[width=3.75in,height=3.5in]{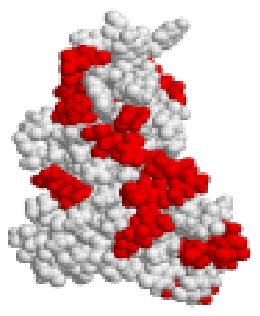}}
        \vspace{2mm}\\
        \subfigure[]{ \label{fig:1tgo_face1}
        \includegraphics[width=3.75in,height=3.5in]{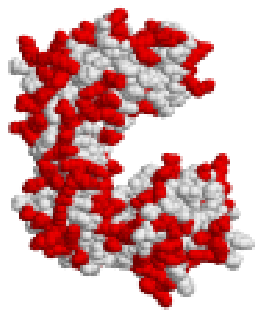}}
\caption{}
\label{fig:surface_charge}
\end{figure}

The availability of a representative structure in these families allows us to map these high probability
sequence correlations (Figs.\ref{fig:P_prions_gaps}) 
and their occurrence in the structure. Mapping onto the NMR structure of the human prion protein
(1QLX) of the positions that are involved in the pairs that correspond to the largest 
$P_{s_k}(i,j)$ (Fig. \ref{fig:1qlx_face1})
shows that these 30 positions (which are mostly found in the 3 helices) are clustered almost entirely on one 
face of the three dimensional prion structure. In the prion family the localization of charged amino acids in the
3D structure is reflected in the specific peaks in $P_{s_k}(i,j)$. A similar
mapping of positions (selected on the same basis as in prions) in DNA-polB on a structure 
from {\em Thermococcous Gorgonarius} (1TGO)
shows that these positions are uniformly distributed on all faces of the structure. If the linear density 
of charges (number of charged residues divided by sequence length) is roughly uniform, we expect  
$P_{s_k}(i,j)$ to decay smoothly without any structure. 
This is the case in DNA-polB family (Figs.\ref{fig:P_prions_gaps}). 
As a result we find that, at the tertiary 
structure level, the charges are roughly uniformly distributed throughout the surface (Fig. \ref{fig:1tgo_face1}).

Could the observation of preferred sequence separation be anticipated 
from sequence entropy calculation alone? To answer this, we calculated
$S(i) = -\sum_{\alpha =1}^{N_{class}} p_{\alpha}(i) ln p_{\alpha}(i)$ using four classes ($N_{class} = 4$) of
amino acids (positively and negatively charged, polar, hydrophobic) and where $p_{\alpha}(i)$ is the probability 
of observing 
amino acid $\alpha$ at site $i$ in a MSA ($p_{\alpha}(i) = \frac{1}{4}$ from random considerations). We
only discuss the prion family because in this family many positions are strongly conserved. There
are a total of 35 charged residues in hPrP(23-253) out of which 14 are perfectly conserved
($S(i) = 0$). In the structured region of hPrP only 6 out of the 14 appear in pairs with high degree of correlation.
Thus, there is no obvious relationship between conservation of individual charged residues and their
covariation as measured by SCE.\\

{\bf Link between number of significant mixed runs and SCE:}
The distribution of $R_{run}(CH)$ for the protein families in the three classes
({\bf HC}, {\bf MC}, {\bf LC}) shows clear 
differences between them (data not shown). In proteins belonging to the
{\bf HC} class (104 members) there is a long tail in the distribution of $R_{run}(CH)$ which implies
that a large number of mixed (CH) runs are found. In the {\bf HC} class we find that a 
significant number of proteins have $R_{run}(CH) >$ 3 whereas the maximum value is 
$R_{run}(CH) <$ 3 for protein families belonging to the {\bf MC} class (397 members). In the {\bf LC}
class (338 members) the number of significant mixed (CH) runs hardly exceeds 2. Overall, about
62\% of families in {\bf HC} class have $R_{run}(CH) >$ 1. On the other hand only 43\% of the
families in the {\bf MC} class have $R_{run}(CH) >$ 1 whereas only about 10\% of families 
in the {\bf LC} class have $R_{run}(CH) >$ 1.

The results for the distribution of $R_{run}(CP)$ show even more dramatic differences
between the three classes. About 21\% of families in the {\bf HC}
class have $R_{run}(CP) >$ 1, whereas only about 2\% of the {\bf MC} class families have multiple
($>$ 1) significant mixed (CP) runs. Among the proteins in the {\bf LC} class we do not find
any protein family with $R_{run}(CP) >$ 1. The percentage of families with either
$R_{run}(CH) >$ 1 or $R_{run}(CP) >$ 1 is 69\%,  44\%, and 10\% in the {\bf HC}, {\bf MC}, and {\bf LC} 
class respectively. These results show that there is a significant correlation between the number of mixed
charged runs and the SCE.

Examples of families with $R_{run}(CH) >$ 1 are aldedh (Aldehyde dehydrogenase, allergens), 
Basic (myogenic Basic domain, DNA binding with bHLH motif),  Bet\_v\_I (pathogenesis-related protein Bet v I
family, allergens), endotoxin (bacteria toxins),
bZIP\_MAF (bZIP MAF transcription factor, cancer-related), Granin (cancer-related), Myc\_N\_term (cancer-related), 
prions, Arena\_nucleocapsid, 
Bunya\_RdRp (Bunyavirus RNA-dependent RNA polymerase), Corona\_nucleocapsid, 
DNA\_polB, Flu\_PB1 (Influenza virus RNA-dependent RNA polymerase subunit PB1), Hanta\_nucleocapsid, 
Paramyxo\_ncap (Paramyxovirus nucleocapsid protein), RHD (cancer-related), 
Tropomyosin (allergens), TTL (breast cancer related),
GroEL, HSP90, P53 and actin. $R_{run}(CP) >$ 1 is found in Androgen\_recep (Kennedy disease),
HSP90, HCV\_capsid (Hepatits C virus nucleocapsid protein), Granin, Myc\_N\_term, P53, Corona\_nucleocapsid, 
Astro\_capsid (Astro virus nucleocapsid precursor), Arte\_nucleocapsid (Arteri virus nucleocapsid protein), 
and Paramyxo\_ncapsid. 

We note that, in general, $R_{run}(CP) >$ 1 occurs only in families of proteins associated with diseases, while
$R_{run}(CH) >$ 1 is found in both families of normal proteins and of proteins associated with diseases.
Even more interestingly, this analysis reveals differences between disease-related proteins
which might be the reflection of the corresponding disease mechanism: proteins found in allergens and toxins and 
the majority of proteins related to cancer have large $R_{run}(CH)$ values and small 
$R_{run}(CP)$ values, while the protein related to Kennedy disease and some of the viral nucleocapsid
proteins have large $R_{run}(CP)$ values and small $R_{run}(CH)$ values. 

A summary of the major findings is given in Table \ref{table:results_all_fam}. There are a number of
lessons that can be gleaned from our findings: (1) Only about 7\% of all normal proteins
have high correlation among charged residues compared to 25\% of all proteins. However, among the 41
``normal'' protein families (that are in the {\bf HC} class) 28 (68\%) have at least one significant
(CH) or (CP) run. (2) The largest percentage of proteins in the {\bf HC} class is from viral
nucleocapsid protein families. These proteins, which are involved in transcription, also have a number
of significant (CH) or (CP) runs. (3) The families of proteins that bind to nucleic acids
in the {\bf HC} class have the highest percentage of combined (CH) and (CP) runs. Taken together
these results show that, for all families, relatively low values of SCE are linked to the number of
significant (CH) and/or (CP) runs.\\

{\bf \large Discussion}\\

{\bf Comparison with other methods for extraction of sequence correlations:}
In the {\bf Methods} section we noted that a quenched average over the sequences in the MSA  can
reveal novel correlations between residues (Eqn.(\ref{eqn:Pk_ij})). 
To ascertain if similar inferences can be drawn using other methods we 
performed a Mutual Information Function (MIF) (Li, 1990) analysis of pairs of charged residues
in the various protein families. We first determined the probability to find a charged residue (type C1) at a 
given position i in a MSA,
\begin{equation}
P_i(C1) = \frac{\sum _{p=1}^{n_{seq}} \delta([i] - C1)}{\sum _{p=1}^{n_{seq}} 1}
\label{eqn:Pi_MIF}
\end{equation}
where $\delta([i] - C1)$ is unity if the amino acid at position $i$ is type $C1$ and is zero otherwise.
The probability of finding a pair of charged residues (C1 and C2) at two
positions (i and j) in the MSA
\begin{equation}
P_{ij}(C1,C2) = \frac{\sum _{p=1}^{n_{seq}} \delta([i] - C1) \delta([j] - C2)}{\sum _{p=1}^{n_{seq}} 1}
\label{eqn:Pij_MIF}
\end{equation}
The MIF is
\begin{equation}
MIF(C1,C2) = \sum _{i = 1}^{L_{FAM}-1} \sum _{j=(i+1)}^{L_{FAM}} P_{ij}(C1,C2) ln(\frac{P_{ij}(C1,C2)}{P_i(C1) P_j(C2)})
\label{eqn:MIF}
\end{equation}
Just as before, we need to factor out the effects of the length of each MSA, so we measured the quantity:
\begin{equation}
MIF^{*}(C1,C2) = \frac{MIF(C1,C2)}{ln(L_{FAM})}
\label{MIF_final}
\end{equation} 
The $MIF^{*}(C1,C2)$ values for all 839 families are much smaller than the maximum possible value
($MIF_{max}(C1,C2)$ = 4.4, because $MIF_{max}(C1,C2) = ln(\frac{1}{P(C1) P(C2)})$ and $P(C1)$ = 0.11). 
The lack of any discernible structure (which is indicative of correlations among the chosen residues) 
in the $MIF^{*}(C1,C2)$ values is likely the result of pre-averaging 
over the sequences in the MSA. Our method, which does not use preaveraging (see Eqn.(\ref{eqn:Pk_ij})), is therefore 
able to capture correlations that are not easily detected by the MIF approach. 
It is worth noting that MIF is suitable for many practical applications. A combination of methods might be
required to obtain correlations between residues using sequence information alone.\\

{\bf Plausible functional link among some families in the HC class:}
The class of proteins that gives the most clear and consistent signal (relatively low values of SCE and multiple
significant (CH) and (CP) runs) is that associated with disease-related proteins like prions, viruses and P53.
Sequence-level correlations could be the result of the constraints imposed
on the evolution of the protein by its function in the cell. 
Using these observations we tentatively {\em hypothesize that protein families with high degree of charge
correlations may have somewhat similar functions}.
If this hypothesis is valid, we can surmise that there may be 
some level of similarity between the actions of prions and those of nucleocapsid viral proteins.
Similarly, the functions of prions and P53 may be somewhat related.
The function of prions is not known, but
those of P53 and nucleocapsid proteins are: they both bind DNA with nucleocapsid proteins playing a vital 
role in the transcription and replication of viral DNA/RNA. Our hypothesis would suggest that the cellular 
form of prions can also bind nucleic acids. Because prions resemble the HCV 
nucleocapsid proteins even at the level of individual $P_{s_k}(i,j)$ 
(Figs. \ref{fig:P_prions_gaps}) we propose that the function of prions could be similar
to that of nucleocapsid proteins. 

There is experimental support to our inference that the functions of prions and nucleocapsid proteins
may be similar. A series of studies
(Sklaviadis et al., 1993; Cordeiro et al., 2001; Gabus et al., 2001a; Gabus et al., 2001b; Moscardini et al., 2002) 
have shown that the 
prion protein has DNA strand transfer properties similar to viral nucleocapsid proteins. 
It has been postulated that in prions an unknown cofactor
(``Protein X'') facilitates the dramatic conformational transition from the predominantly 
$\alpha$-helical structure to a state rich in $\beta$-sheet. 
DNA strands could play the role of ``protein X'' in the conformational
transition.\\

{\bf Number of significant (CH)/(CP) runs and propensity for scrapie formation in prions are linked:}
Recent sequence and structural analysis (Kallberg et al., 2001; Dima and Thirumalai, 2002) 
has suggested that elements of secondary structure in mammalian
prions are frustrated. By frustration we mean that the secondary structure elements in the normal
cellular form
could be better accommodated by an alternative conformation. Avian prion
sequences are not frustrated (Dima and Thirumalai, 2002) which explains the lack of observation of 
the scrapie form in these species.
These findings are further corroborated here by the variations in the significant (CH) and (CP) runs
(Table \ref{table:Composition_vs_runs_prions}) between these species.
Despite the lack of significant differences (see Table \ref{table:Composition_vs_runs_prions}) 
in the amino acid composition
(especially in charged and polar residues) among the various species, the chicken prion sequence as well as 
of the other avian species does not have
significant mixed charged-hydrophobic/polar runs. However, (CH) and/or (CP) runs 
appear in all mammalian prions
which are known to be associated with prion diseases. 
The absence of such runs might be one of the reasons for the lack of formation
of the scrapie form of prions in avian species.\\

Despite the deleterious effects of sequence correlations among charged residues in proteins associated with
diseases, our findings suggest 
that charged amino acids must play an important role in the
functions of these proteins. Viruses (like HIV and hepatitis C) have high degree of
sequence correlations between charges. Blocking the repertoire of 
charges in viruses might impair of their capacity to induce and 
promote the associated disease. As a rule, protein families  
with high degree of sequence correlations also have a
significant number of mixed charged-hydrophobic/polar runs. When charged residues appear correlated in a sequence then
they are likely to be distributed in patches in the three dimensional structure. 

Given the potential link between high degree of charge correlation and disease it is not clear why these
proteins have not evolved with more benign characteristics. Perhaps, there are functional demands on
this class of proteins that require multiple runs and significant charge correlations.
The high degree of charge correlation and the presence of mixed (CH) and/or (CP) runs
might be indicative of their role in protein-protein interactions or binding to DNA or RNA.
It is also likely that the charge distributions may also be
important in avoiding aggregation. As a corollary, we find that the majority of
``normal'' proteins exhibit only moderate or weak sequence correlation between charges. The identification 
of correlated charged pairs in various families suggests that
mutation of these residues can compromise their function. This prediction is amenable to
experimental tests.\\

{\bf Importance of charged residues in kinetics of fibrillization:} The factors that affect the amyloid 
deposition rates have not been fully elucidated. Only recently a systematic physical basis that relates
sequence characteristics in disease-related proteins and amyloid formation has been explored 
(Chiti et al, 2003). This study shows that the overall charge states greatly affect fibrillization kinetics.
The deposition rate decreases as the overall charge of the disease related proteins increases. Similarly, we had
argued (Thirumalai et al., 2003)
both in prion proteins and A$\beta$ peptides that the overall charge is relevant for polymerization.
For example, the A$\beta_{16-22}$ peptide with the sequence KLVFFAE is a significant
(CH) run. This peptide has been shown by using solid state NMR measurements to readily aggregate
into amyloid fibrils organized into antiparallel $\beta$-sheets (Balbach et al., 2000). Extensive Molecular Dynamics
simulations in explicit solvent probing the dynamics of the assembly process for 
three A$\beta_{16-22}$ peptides (Klimov \& Thirumalai, 2003) revealed
that electrostatic and hydrophobic interactions play different roles in the formation of the
antiparallel $\beta$-sheet: the electrostatic interactions play a crucial role in the orientation
of the peptides in the oligomer, while the hydrophobic interactions bind the peptides together. 
As a result, mutations at either the C-term end from a negatively charged residue to polar residues 
(E22G and E22Q) or at
the middle hdrophobic positions (L17S/F19S/F20S)  reduce considerably the stability of the oligomer.     
These studies provide additional support for our prediction that charged residues clustered into
significant (CH) runs play an important part in the dynamics of protein aggregation.
The prediction, based solely on bioinformatic analysis,
that correlated mutations can inhibit amyloid formation can be experimentally tested.\\

{\bf Acknowledgments:}
We have benefited from critical discussions with Dmitri K. Klimov and Changbong Hyeon. 
This work was supported in part by a
grant from the National Institutes of Health (IR01 NS41356-01).

\begin{sidewaystable}
\centering
\caption{Sequence correlation entropy for charged residues in protein families from the {\bf HC} class}
\renewcommand{\thefootnote}{\thempfootnote}
\begin{tabularx}{175mm}{XXXXX} 
\hline
 Family(database)  & SI\footnote{The sequence identity in the family.} & (+,+)\footnote{Sequence correlation entropy, expressed as a percentage (see Eq.(\ref{eqn:ratio_Sij})) for (+,+) pairs.}  & (+,-)\footnote{Same as b, except it is for (+,-) pairs.}   & (-,-)\footnote{Same as b, except it is for (-,-) pairs.}   \\ 
\hline
 Prions            & 83 & 48  & 34   & 42 \\
 Prions(Blast)     & 72 & 45  & 33   & 42 \\
 A$\beta$(Blast)   & 82 & 31  & 33   & 40 \\
 Tau(Blast)        & 55 & 49  & 37   & 49 \\
 GroEL(Blast)      & 54 & 50  & 37   & 46 \\
 A2M               & 51 & 50  & 38   & 48 \\
 Adeno-hexon       & 51 & 47  & 36   & 45 \\
 Cytochrome-b-C    & 74 & 43  & 34   & 31 \\ 
 DNA-polB          & 27 & 50  & 39   & 53 \\
 Gag-p17           & 69 & 52  & 38   & 43 \\
 HCV-capsid        & 90 & 39  & 37   & 41 \\
 HH-signal         & 84 & 42  & 29   & 39 \\
 IF3               & 45 & 49  & 37   & 52 \\
 Lipoprotein-1     & 71 & 47  & 33   & 46 \\
 P53               & 63 & 44  & 37   & 46 \\
 PAL               & 62 & 48  & 36   & 47 \\
 recA              & 67 & 51  & 39   & 47 \\
 Rubisco-large     & 85 & 48  & 38   & 50 \\
 Vpu               & 62 & 45  & 32   & 49 \\
\hline
\end{tabularx}
\label{table:correl_charged_gaps}
\end{sidewaystable}

\begin{sidewaystable}
\centering
\caption{Link between sequence correlations and the number of mixed runs for the 839 families}
\vspace{1cm}
\renewcommand{\thefootnote}{\thempfootnote}
\begin{tabularx}{200mm}{XXXXXXX} 
\hline
Family\footnote{The first column describes the functions of the families in the data set
(see text for an explanation).  The 450 members in the ``Normal'' category
exclude the ``NA'' and ``HSP'' families.} & HC\footnote{Percentages of protein families in the Highly Correlated ({\bf HC}), Moderately Correlated ({\bf MC}), and Loosely Correlated ({\bf LC}) classes. The largest and the smallest
percentages in each column are given in bold and in italics respectively.} & MC\footnotemark[\value{mpfootnote}] & LC\footnotemark[\value{mpfootnote}] & (CH)\footnote{ The percentage of families, among the {\bf HC} class, that have at least one significant mixed
charged-hydrophobic run.} & (CP)\footnote{Same as (e), except it is for mixed charged-polar runs.} & run\footnote{Percentage of proteins, in the {\bf HC} class, that have either one (or more)
significant (CH) or one (or more) significant (CP) run(s).} \\
\hline

All & 25 & 43 & 32 & 62 & 21 & 69 \\

Normal & {\em 7} & 44 & 49 & 63 & 21 & 68 \\

HSP & 10 & 38 & {\bf 52} & 100 & 50 & 100 \\

NA & 15 & 48 & 37 & {\bf 77} & 23 & {\bf 82} \\

Disease & 23 & {\em 35} & 42 & 67 & {\bf 45} & 78 \\

Viruses & 34 & {\bf 54} & 12 & 55 & 25 & {\em 65} \\

Capsid & {\bf 58} & 39 & {\em 3} & {\em 53} & 40 & 67\\ 
\hline
\end{tabularx}
\label{table:results_all_fam}
\end{sidewaystable}
\[
\]

\begin{sidewaystable}
\centering
\caption{Amino acid composition and number of mixed runs in prion proteins}
\vspace{1cm}
\renewcommand{\thefootnote}{\thempfootnote}
\begin{tabularx}{200mm}{XXXXXXX}
\hline
Species & H\footnote{Percentage of hydrophobic residues.} & C\footnote{Percentage of charged residues.} & P\footnote{Percentage of polar residues.} & L\footnote{Number of amino acids.} & N$_{run}$(CH)\footnote{Number of significant (CH) runs.} & N$_{run}$(CP)\footnote{Number of significant (CP) runs.} \\
\hline
Human & 28 & 17 & 55 & 208 & 2 & 0 \\

Possum & 34 & 15 & 51 & 235 & 2 & 1 \\

Turtle & 33 & 16 & 51 & 246 & 0 & 2 \\

Chicken & 36 & 15 & 49 & 249 & 0 & 0 \\
\hline
\end{tabularx}
\label{table:Composition_vs_runs_prions}

\end{sidewaystable}
\[
\]

\newpage

\begin{center}
{\bf Figure Captions}
\end{center}

Fig.\ref{fig:SCE_family_seqid}. (a):  Histograms of $\bar{S}(i,j)$
(Eq.(\ref{eqn:ratio_Sij})) for the 839 families. The top, middle, and bottom panels are for
(-,-), (+,-), and (+,+) respectively.
(b): The distribution of average sequence identity (SI)
(as presented in the Pfam file of the protein family) for the families in
our dataset. Each family was classified according to its class ({\bf HC}, 
{\bf MC} or {\bf LC}) as described in the text. 

Fig.\ref{fig:PCA_on_SCE}. (a): The plot of the first and fourth eigenvectors of the matrix of
euclidean distances between pairs of families (see text for details). (b): The plot of the second and
fourth eigenvectors of the distance matrix. The three distinct regions in these graphs allow the
classification of families in three classes: {\bf HC}, {\bf MC}, and {\bf LC} (see text for details). 

Fig.\ref{fig:P_prions_gaps}: The probability, $P_{s_k}(i,j)$, of finding residues i and j at separation
$s_k$ as a function of $ln (s_k)$ for the prion, HCV capsid and DNA-polB families.
(a) (i,j) = (+,+), (b) (i,j) = (+,-), (c) (i,j) = (-,-). Except for DNA-polB, the presence of peaks in
$P_{s_k}(i,j)$ suggests enhanced sequence correlation at specific separations of charges.

Fig.\ref{fig:surface_charge}: Mapping of highly correlated, as measured by  $P_{s_k}(i,j)$
(see Fig.(\ref{fig:P_prions_gaps})), charged residues onto the three dimensional structure.
(a) Space filling representation of one face of the human prion structure (1QLX). Residues
appearing in pairs corresponding to the peaks in $P_{s_k}(i,j)$ (Fig.(\ref{fig:P_prions_gaps})) are
shown in black. Highly correlated charged residues in sequence space are localized in the
three dimensional structure. (b) Similar picture for 1TGO, which is a  
representative structure for DNA-polB. The black shades represent residues that have significant values of
$P_{s_k}(i,j)$ (Fig.(\ref{fig:P_prions_gaps})). Charged residues are uniformly distributed in the 
three dimensional structure.

\newpage

\begin{figure}[htbp]
        \subfigure[]{
        \label{fig:S_pair(C1,C2)_gaps}
        \includegraphics[width=4.0in,height=3.5in]{plot_entropy_all_gaps_C.eps}}
        \vspace{5mm}\\
        \subfigure[]{ \label{fig:Class_family_Seqid}
        \includegraphics[width=4.0in,height=2.75in]{Class_family_Seqid.eps}}
\caption{}
\label{fig:SCE_family_seqid}
\end{figure}

\newpage

\newpage

\begin{figure}[htbp]
  \vspace{-1.5cm}
        \subfigure[]{
        \label{fig:P1_prions_gaps}
        \includegraphics[width=3.25in,height=2.0in]{P1_prion_gap.eps}}
        \vspace{1mm}\\
        \subfigure[]{ \label{fig:P2_prions_gaps}
        \includegraphics[width=3.25in,height=2.0in]{P2_prion_gap.eps}}
	\vspace{1mm}\\
	\subfigure[]{ \label{fig:P3_prions_gaps}
        \includegraphics[width=3.25in,height=2.0in]{P3_prion_gap.eps}}
\caption{}
\label{fig:P_prions_gaps}
\end{figure}

\newpage

\end{document}